# Accumulation and control of spin waves in magnonic dielectric microresonators by a comb of ultrashort laser pulses

A.E. Khramova [1,2], M. Kobecki [3], I.A. Akimov [3,6], I.V. Savochkin [1,2], M.A. Kozhaev [1,4], A.N. Shaposhnikov [5], V.N. Berzhansky [5], A.K. Zvezdin [1,4,7], M. Bayer [3,6], V.I. Belotelov [1,2]

[1] Russian Quantum Center, 45, Skolkovskoye shosse, Moscow, 121353, Russia
[2] Faculty of Physics, Lomonosov Moscow State University, Leninskie Gory, Moscow 119991, Russia
[3] TU Dortmund, Experimentelle Physik 2, D-44221 Dortmund, Germany
[4] Prokhorov General Physics Institute RAS, 38 Vavilov Street, Moscow 119991, Russia
[5] Vernadsky Crimean Federal University, 4 Vernadskogo Prospekt, Simferopol, 295007, Russia
[6] Ioffe Institute, Russian Academy of Sciences, 194021 St. Petersburg, Russia
[7] NTI Center for Quantum Communications, National University of Science and Technology MISiS, Leninsky Prospekt 4, Moscow, 119049, Russia

**ABSTRACT**

Spin waves in magnetic microresonators are at the core of modern magnonics. Here we demonstrate a new method of tunable excitation of different spin wave modes in magnetic microdisks by using a train of laser pulses coming at a repetition rate higher than the decay rate of spin precession. The microdisks are etched in a transparent bismuth iron garnet film and the light pulses influence the spins nonthermally through the inverse Faraday effect. The high repetition rate of the laser stimulus of 10 GHz establishes an interplay between the spin wave resonances in the frequency and momentum domains. As a result, scanning of the focused laser spot near the disk boarder changes interference pattern of the magnons and leads to a resonant dependence of the spin wave amplitude on the external magnetic field. Apart from that, we achieved a switching between volume and surface spin waves by a small variation of the external magnetic field.

**INTRODUCTION**

Currently, research in the field of magnonics is of great importance. The practical application of spin waves is being rapidly developed as an alternative to existing methods of transmitting digital information as well as for use in quantum communications [1,2]. These days, the first experimental developments of magnon logic systems have already been obtained [3,4]. They are based, for example, on the interaction of spin waves with domain walls of various chiralities, which reduces energy dissipation [5]. The development of spin resonators and interferometers for telecommunication devices is also actively underway [6–8]. For most of practical magnonics implementations, magnetic dielectrics play a key role as their Gilbert damping is much lower than for magnetic metals. Among them, yttrium iron garnet thin films demonstrate a record performance [9,10]. For example, yttrium iron garnet spheres of millimeter size were recently coupled to superconducting qubits in a microwave cavity to provide control and quantum data storage [11].

Usually, spin waves are excited by microwave antennas [12–14]. This method has strong constrictions in tunability. For example, to change the position of the region where the spin waves are excited in a sample, a mechanical movement is necessary which limits operation rates and introduces additional noise. Apart from that, microwave antennas influence on relatively large part of a magnetic sample which worsens localization of the spin wave source. In this respect, optical means involving femtosecond laser pulses hold a large promise as they allow one to solve most of these obstacles [15–23]. Ultrashort laser pulses disturb the equilibrium magnetic state of the sample either via non-thermal effects like the inverse Faraday and Cotton-Mouton effect, or by the effect of the photoinduced magnetic anisotropy, or through the thermal effect of ultrafast demagnetization [24]. The former are mostly prominent for magnetic dielectrics, for example, bismuth iron garnets [25] as they possess low optical absorption. The inverse Faraday effect arises due to the stimulated Raman scattering of light by magnons which can be understood in terms of the circularly polarized light creating an effective magnetic field along the light wavevector. A focused laser beam influences only the spins within the laser spot and during the pulse propagation through the sample. However, it is sufficient to trigger the magnetization precession. On a picosecond time scale, adjacent spins get also involved through the magnetic dipole-dipole and exchange interactions and the magnetic oscillations propagate away from the illuminated area in the form of spin waves.

Among the drawbacks of the optical method are the relatively broad spectrum of the launched spin waves and their small amplitudes. Usually, laser pulses excite the spin dynamics in a single pulse regime, where the magnetization precession started by one pump pulse completely decays before the arrival of the next pulse. Excitation in the multiple pulse regime where the time interval between pulses is shorter than the decay time is much more advantageous. Recently, it was demonstrated that the periodic laser impact on spins provides a kind of accumulation of the spin waves whose spectrum and directionality can be modified by a slight variation of the applied magnetic field or the pulse repetition rate [16]. Moreover, in this case the spin waves are excited in a frequency range much narrower than for single pulse excitation and with amplitudes prominently enhanced. Indeed, a train of fs-laser pulses provides a kind of resonance in the frequency domain and thus singles out spin wave frequencies that are multiples to the pulse rate. Taking into account that self-mode-locked solid state lasers are capable of generating optical pulses with the repetition rates up to several tens of GHz such approach enables optical generation of high frequency sub-THz magnons.

On the other hand, magnetic structures like magnonic waveguides, microresonators and magnonic crystals have become crucial elements of modern magnonics[26–34]. Multiple spin wave reflections from the boundaries and interfaces lead to resonances in the momentum domain. As a result, complex interference patterns are formed and additional resonances in the spin wave spectra appear. Magnonic microstructures are necessary for expanding the capabilities of magnonic devices as the micro- and nanostructuring enrich magnonic spectra and open new doors for their engineering. Moreover, magnonic microspheres and microdisks play the role of cavities for both magnons and photons and therefore are currently used for quantum magnonics platforms [11].

Here we combine the approaches of periodic laser impact and spatial confinement of spin waves to investigate the interference between spin wave resonances in the frequency and momentum domains. In particular, we study magnetostatic spin waves in the iron-garnet microdiscs of different diameters, excited by a sequence of ultrashort laser pulses. The most interesting situation appears for relatively small disk diameters and laser comb periods so that the disk diameter doesn't exceed the propagation distance of magnons and the pulse interval is kept shorter than the magnon decay rate. In this case both the spatial confinement and the periodic laser impact provide accumulation of magnons: the spin waves are closed in an area of the microdisk and their energy is permanently filled up from coming laser pulses. On the other side, presence of the microresonator and the laser comb singles out standing spin modes and make their frequency to be a multiple of the pulse repetition rate. If an external magnetic field is swept, which tunes frequency of the spin modes, it allows to control amplitude of the modes. Moreover, in such regime it becomes possible to easily switch between different types of spin wave modes. Additional control of spin waves becomes possible due to the scanning the spot of the periodic laser pulses over the microdisk area since it changes interference conditions for the magnons reflecting from the disk boarders.

**RESULTS AND DISCUSSION**

The magnetic sample of the bismuth iron garnet microdisks and configuration of the experiment are presented in Fig. 1a. The spin dynamics in the microdisks is excited by a comb of laser pulses with a period of 100 ps and it is observed by the Faraday effect of the probe beam (see Methods for details).

We start the detailed studies from the 300-µm microdisk and illuminate its central part by the laser spot of a much smaller diameter. Therefore, the influence of the disk boundaries can be neglected and the disk can be considered approximately as an infinite smooth magnetic film. The observed spin dynamics is a result of the excited magnetostatic spin waves. This is confirmed when the position of the pump and probe beam spots are separated, i.e. the observation area is moved away from the illuminated one: the oscillations are still observed but their amplitude decays with the distance between pump and probe spots (Fig. 1(b)). The exponential decay length of the spin waves was approximately 9 μm (Fig S2). The signal is hardly visible 20 μm apart from the excitation spot. Actually, the observed short-range propagation of the spin waves results mostly from the dephasing effect appearing due to the relatively broad spectrum of the spin waves launched by the tightly focused laser pulses.

The dependence of the precession phase on the distance between the pump and probe spots measured at different magnetic fields allows one to derive the dispersion of the spin wave wavelength on the magnetic field[9] (Fig. 1(c)). Apart from that, the spin wave dispersion $\omega(k)$ can be extracted (Fig. 1(d)) which corresponds to the backward volume magnetostatic spin waves (BVMSW) since their frequency decreases for larger wavenumbers.

Figure 1(b) also indicates that the amplitude of the magnetization precession depends strongly on the magnetic field and has a resonant shape. The resonant amplification of the precession amplitude appears due to the synchronization of the instant pump pulses driving the magnetization with the time interval $T$ and the magnetization precession at frequency $\omega$ [16].

Circularly polarized light influences the magnetization in magnetic dielectrics mainly through the inverse Faraday effect which can be described in terms of an effective magnetic field $H_F$, acting on the spins during the pulse propagation through the sample [35]. Since the pulse duration $\Delta t$ is sufficiently small compared to the laser pulse repetition period ($\Delta t \ll T$), the field of the inverse Faraday effect can be represented as a series of $\delta$-functions: $H_F(t) = h\Delta t \sum_{l=0}^{+\infty} \delta(t-lT)$, where $h$ is the field amplitude, and $l$ is an integer. Solving the Landau-Lifshitz-Gilbert equation written in spherical coordinates for the periodic process of period $T$ gives the spin precession in the form of $\theta(t) = \theta_0 \sin(\omega t + \xi) e^{-t/\tau_0}$ in the interval $lT < t < (l+1)T$. Here $\omega = \omega_0 \sqrt{1-(2\pi\nu_0\tau_0)^{-1}} \approx \omega_0$, $\omega_0 = \sqrt{\omega_H(\omega_H + \omega_d - \omega_a)}$, $\tau_0 = \frac{2}{\alpha(2\omega_H + \omega_d - \omega_a)}$, $\omega_H = \gamma H$, $\omega_d = \gamma H_d$, $\omega_a = \frac{2\gamma K_u}{M}$, $H_d = 4\pi M_s$ is demagnetizing field which arises when the sample is magnetized in the film plane. $\theta_0$ and $\xi$ are the amplitude and initial phase of the oscillations, respectively[15]. The precession amplitude is given by

$$\theta_0 = \frac{\gamma^2 H h \Delta t}{\omega}\left(1 - 2e^{-T/\tau_0}\cos\omega T + e^{-2T/\tau_0}\right)^{-\frac{1}{2}} \quad (1)$$

Therefore, the amplitude reaches maximum values at $\omega = 2\pi m/T \equiv \omega_r$, where $m$ is an integer and the dependence $\theta(H)$ should have a resonance shape. Here we deal with pulses coming at a high repetition rate, so that $T/\tau_0 \ll 1$ and the resonances are relatively narrow, i.e. $\Delta\omega T \ll 1$, where $\Delta\omega$ is the resonance width. In this case, Eq.(1) in the vicinity of the resonance simplifies to a Lorentzian shape:

$$\theta_0 = \frac{\gamma^2 H_{r0} h \Delta t}{2\pi m \sqrt{(\omega-\omega_r)^2 + \tau_0^2}} \quad (2)$$

where

$$H_{r0} = \frac{2\pi m}{\gamma T} - \frac{1}{2}\left(H_d - \frac{2K_u}{M}\right) \quad (3)$$

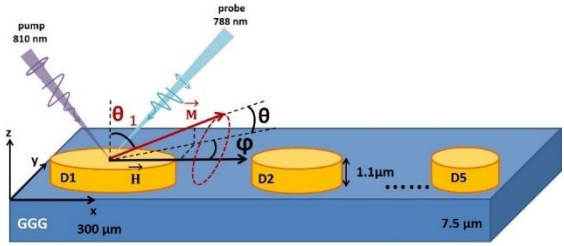
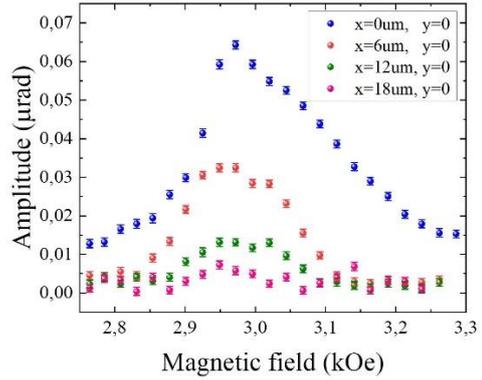
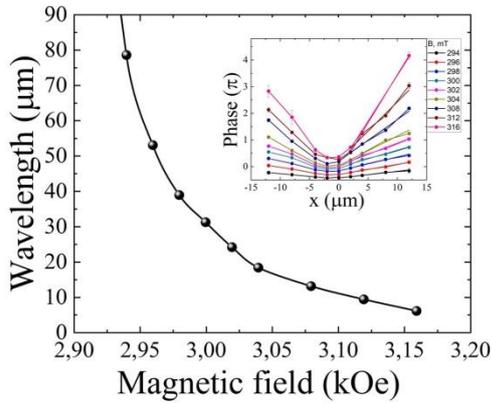
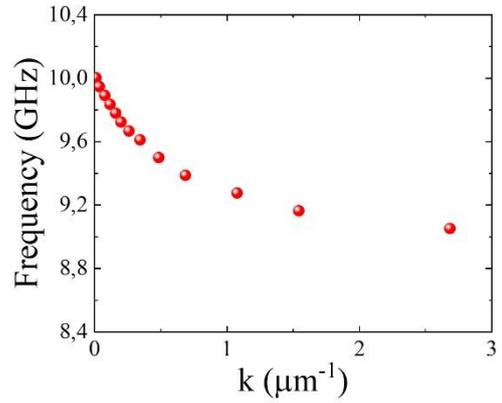
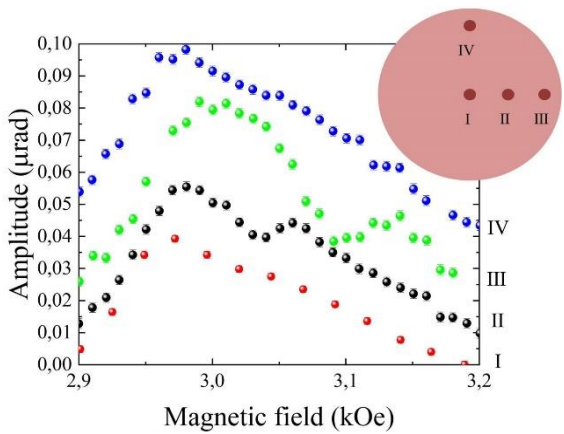
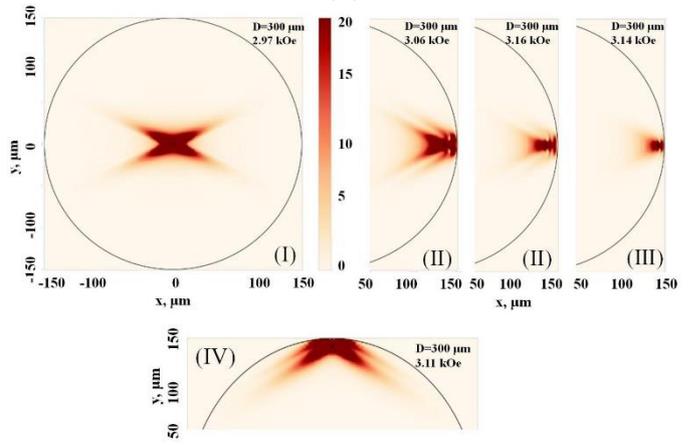

(a)             (b)

(c)             (d)

(e)             (f)

**Figure 1.** Experimental configuration (a) and the spin waves excited in the 300µm disk (b-e).(fig.1(a) was drawn using power point software) (b) Dependence of the spin wave amplitude on the external magnetic field for different positions of the probe beam relative to the pump one. The pump beam is located in the center of the disk(c) Dependence of the spin waves wavelength on the external magnetic field derived from the behavior of the spin wave phase along the x-axis (see inset) measured for different magnetic fields. (d) Experimentally evaluated spin wave dispersion corresponding to backward volume magnetostatic spin waves (BVMSW). (e) Dependence of the spin waves amplitude on the applied magnetic field at four different positions of the pump and probe beams. (f) Simulated distribution of spin wave amplitude at four points indicated in (e). The inset shows these four positions on the microdisk. The pump fluence is 5,2 µJ/cm$^2$

is the external magnetic field of the resonance in the frequency domain. These expressions describe the observed spin dynamics in the limit of small wavenumbers $k$ of the spin waves ($kR \ll 1$, where $R$ is the size

of the monitored area)

The magnetic field strength can be adjusted such that there is only one precession cycle between two consecutive pulses (*m*=1), and the damping of the precession is hardly visible. This is the case of our experiments (see Methods, Fig. 3(a)). Tuning the external magnetic field varies the precession frequency, allowing one to finely synchronize it with the laser pulse train repetition rate. The highest amplitude of the precession should be achieved when its period coincides with the time interval between the pulses, i.e. at $H_{r1}$ =2.96 kOe. This corresponds well with the position of the maximum in the experimental dependence of $\theta_0(H)$ in Fig. 1(b), blue circles, at 2.98 kOe.

Some deviations and the nonsymmetric shape of the resonance $\theta_0(H)$ are due to the excitation of the BVMSW with non-negligible *k*. Actually, the focused laser beam of radius *r* excites BVMSW with wavevectors *k*<2/*r*. Due to the BVMSW dispersion $\omega(k)$ the spectrum of the excited spin waves is broadened: $\omega(1/r) < \omega < \omega_0$. Therefore, the resonance conditions for the different spin wave harmonics occur at slightly different magnetic fields broadening the resonance of $\theta_0(H)$ and shifting it slightly.

The influence of the microdisk boundaries can be studied for the 300-μm structure as the spin wave dispersion in the disk is known (Fig. 1(c)). Let us compare four positions of the pump and probe spot in the disk area (Fig. 1(e)): (I) at the center of the disk (red circles); (II) at a distance *L*=15 μm apart from the edge of the disk along the direction of the magnetic field, i.e. along the *x* axis (black circles); (III) at a distance L= 7 μm apart from the edge of the disk along the x axis (green circles), and (IV) at a distance *L*= 7 μm apart from the edge of the disk in a direction perpendicular to the magnetic field, i.e. along the *y* axis (blue circles).

For the illumination at the microdisk center (I) the amplitude dependence $\theta_0(H)$ has a single peak at $H_{r1}$=2.98 kOe, while for the other positions there are additional peaks in the $\theta_0(H)$ dependence. The main peak at around 2.98 kOe is due to the synchronization of the magnetization oscillations and pump pulses, as we discussed earlier. However, the most important feature of the $\theta_0(H)$ dependence is the second peak (see Fig. 1(e), black and green circles). Its position shifts to higher fields for the light spot moving closer to the microdisk edge. For the *L*=15 μm separation from the edges (point (II)) the peak appears at the magnetic field of 3.06 kOe, while for *L*=7 μm (point (III)) - at 3.14 kOe. Apart from that, for the point (II) there is also a minor peak at 3.16 kOe and for the point (III) there is a feature at 3.04 kOe, that broadens and shifts the main resonance to higher magnetic fields.

The observed behavior of the spin wave amplitude hints at interference between spin waves directly excited by the pump pulses and reflected from the microdisk edges. In detail, the spin waves excited at the center of the 300-μm disk don't reach the disk boundaries since their propagation length is much smaller of the order of a few tens of microns. As a result, no interference is possible (see Fig. 1 (f)), simulations for the point (I)) performed by the micromagnetic modeling program Mu-Max with the cell dimensions of 0.25 × 0.25 × 1.1μm$^3$. The Hilbert dissipation constant in the simulation was taken as $\alpha = 4 \cdot 10^{-3}$. However, if the spin waves are launched a few microns apart from the microdisk edge, namely at a distance *L*, then the interference becomes possible and should be most efficient for the spin wave wavelengths $\lambda \sim \frac{2L}{n}$ where *n* is an integer, so that the phase difference between the spin waves directly excited by the pump and the one reflected from the edges is close to 0 or 2π.

At $H_{r2}$=3.06 kOe (the resonance field at the point (II)) the average wavelength of the BWMSW is $\lambda$ =15 μm, while at $H_{r2}$=3.14 kOe (the resonance field at the point (III)) $\lambda$ =7 μm (Fig. 1(c)). In both cases it nicely agrees with the condition $L \sim \frac{\lambda n}{2}$ for *n*=2 and with simulation results (Fig. 1 (f), simulations for the points II and III).

Thus, the additional peaks are indeed a result of BWMSW interference. It should be noted that BWMSW propagate mostly along the magnetic field, i.e. along x-axis. This makes the reflection from the boundary near the point (IV) negligibly small and could not cause any resonances at the point (IV). However, at this point there is still some marginally seen feature at 3.11 kOe (Fig. 1(e), blue circles). It indicates interference of some spin waves propagating perpendicular to the external magnetic field. The oscillation distribution also points out the interference (Fig. 1f, (IV)) Such spin waves are known as Damon-Eshbach or magnetostatic surface spin waves (MSSW). In this case their excitation efficiency is much lower than for BWMSW. Actually, for the pump beams of relatively large diameters MSSWs are launched worse than BWMSWs[36].

As the size of the disks decreases, additional maxima of the precession amplitude appear as well whose origin can be associated with interference. However, in this case, the size of the disk is comparable to the propagation length of the spin wave, and the phenomenon of interference becomes more complicated. The strong influence of the boundaries is manifested in the behavior of the magnetization oscillation amplitude for different disks (Fig. 2(a)). For smaller disks the field $H_{r0}$ starts to increase. For example, for the 20 μm disk $H_{r0}$=3.01 kOe (Fig. 2(a), red circles). The peak shift by 30 Oe is due to the smaller average demagnetizing field $H_d$ of the microdisk with respect to the film. Eq.(3) gives the correct value of $H_{r0}$ if the average demagnetizing field of the 20 μm disk is $H_d$=1.24 kOe.

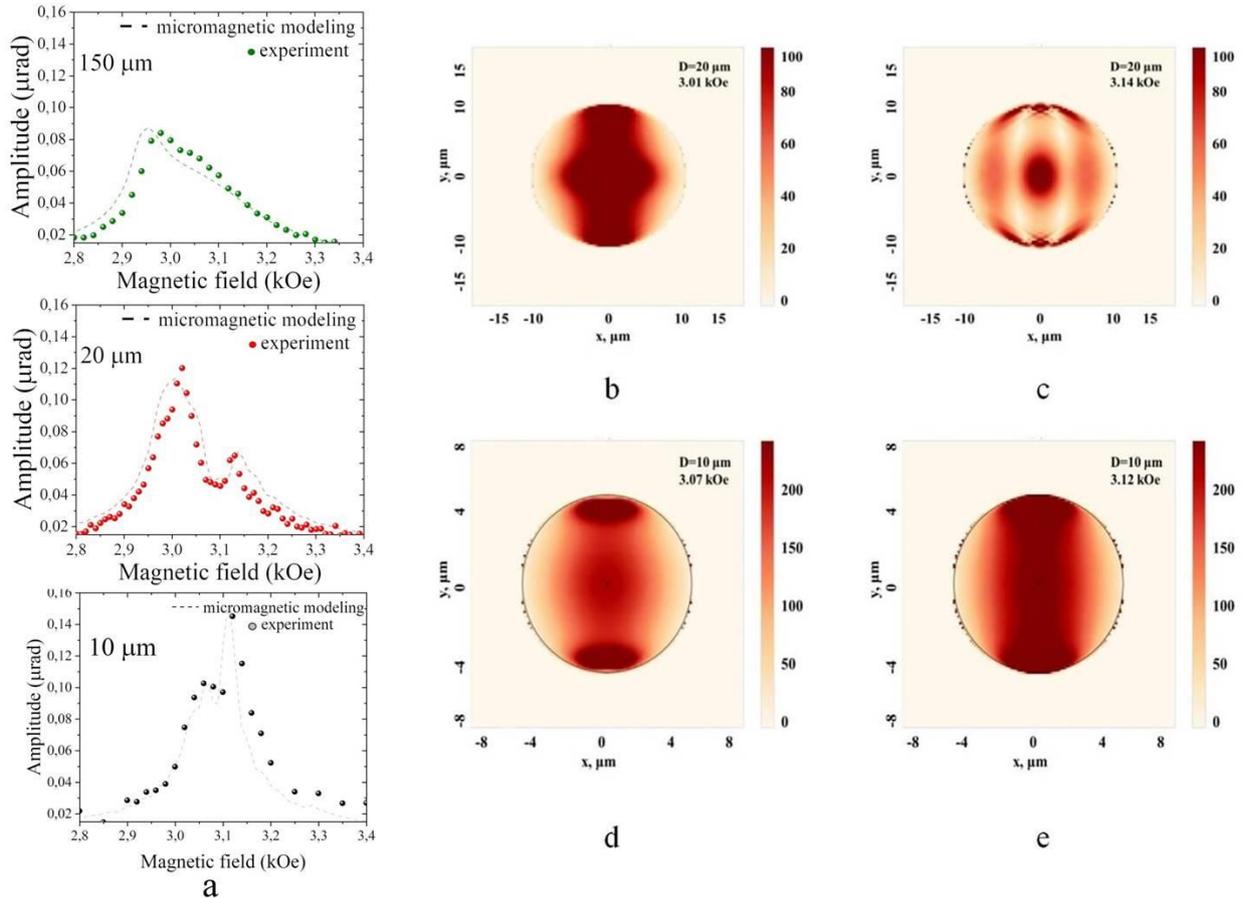

**Figure 2.** (a) Experimental data of the amplitude of the magnetization precession dependence on the magnetic field for disks of 150, 20 and 10 μm diameter (circles), as well as micromagnet modeling results of the magnetization precession amplitude (dashed lines), the energy density is 5.2 μJ/cm$^2$. (b-e) Spatial distribution of the spin wave amplitude for the 20 μm disk at H = 3.01 kOe (b) (the lower field resonance peak), and at H = 3.14 kOe (c) (the higher field resonance peak), for the 10 μm disk at H = 3.07 kOe (d) (the lower field resonance peak), and at H = 3.12 kOe (e) (the higher field resonance peak).

Notably, for the 20 μm disk a second peak appears at $H_{r2}$=3.14 kOe. Its origin can be understood from the calculated spatial distribution of $\theta_0(x,y)$ (Fig. 2(b-e)). Here the best correspondence between experimental and modeling data is achieved for the fully unpinned boundary conditions. Periodic pumping at 10 GHz repetition rate singles out different magnetostatic modes of the disk formed by standing spin waves. As the modes are formed by BVMSWs, the interference takes place along the external magnetic field, i.e. one or several amplitude maxima appear in x-axis direction. Thus, the mode excited at $H_{r0}$=3.01 kOe is the standing spin wave of the first order ($n_x$=1), while the mode at $H_{r2}$=3.13 kOe is of the third order ($n_x$=3).

The amplitude of the spin precession in the 10 μm disk also has two peaks at 3.07 kOe and 3.12 kOe. At a magnetic field of 3.12 kOe the first-order ($n_x$=1) mode of the BMSW is excited (Fig. 2(e)), while origin of the lower field peak at 3.07 kOe should be different since the amplitude distribution doesn't correspond to any

BMSW modes and it demonstrates interference in y-axis direction (Fig. 2(d)).Moreover, it follows from the amplitude distribution, that the modes is also of the first order, therefore its wavelength obeys the same condition: $\lambda=2D$. As resonance magnetic fields for the two modes are different it means that the modes have different dispersion and, therefore, different character. It hints again on the excitation of MSSW as it was observed for the point (IV) in the 300-μm disk. However, this time their resonance is more pronounced. Amplitude distribution in Fig. 2(d) confirms interference along y-axis and MSSW character in this case. Therefore, the lower field resonance peak is due to the first order ($n_y$=1) MSSW mode. Thus, a spatial confinement provides more efficient excitation of MSSWs due to the formation of the standing MSSWs. The resonance is rather sharp which allows to tune between MSSW to BWMSW by varying the external magnetic field by 60 Oe only.

### CONCLUSIONS

To conclude, we have studied the excitation of spin wave modes by a sequence of ultrashort laser pulses impinging on the magnetic microdisks at the ultimately high repetition rate of 10 GHz. The focused laser spot acts as a localized stimulus for launching the spin oscillations which significantly sets it apart from conventional microwave generation means. The high repetition rate leads to spin wave accumulation. Due to the reflection of the spin waves from the disk boundaries the interference and standing mode formation take place. It provides a switching between BVMSW and MSSW by a small variation of the external magnetic field. An additional degree of tunability might be achieved by scanning the laser spot across the microdisk area which provides different interference patterns. The experimental results were obtained with the laser focused to a spot of 7 μm in diameter. If a tighter focusing is used, around 1 μm in diameter, then it might be possible to scan across a smaller microdisk and launch not only odd modes, but even modes as well. This makes the proposed method of excitation of spin waves advantageous for the possible application of magnonics in data processing systems.

### METHODS

The multiple pulse impact on the magnetization in the microdisks initiates a quasi-stationary process of magnetization precession (Fig. 3(a)). The magnetic oscillations excited by a pump pulse don't noticeably decay before the consecutive pulse hits the samples. However, their amplitudes strongly depend on the magnetic field *H* (Fig. 3(b)). Moreover, for some disks this dependence has a double peak structure (Fig. 3(b), red circles).

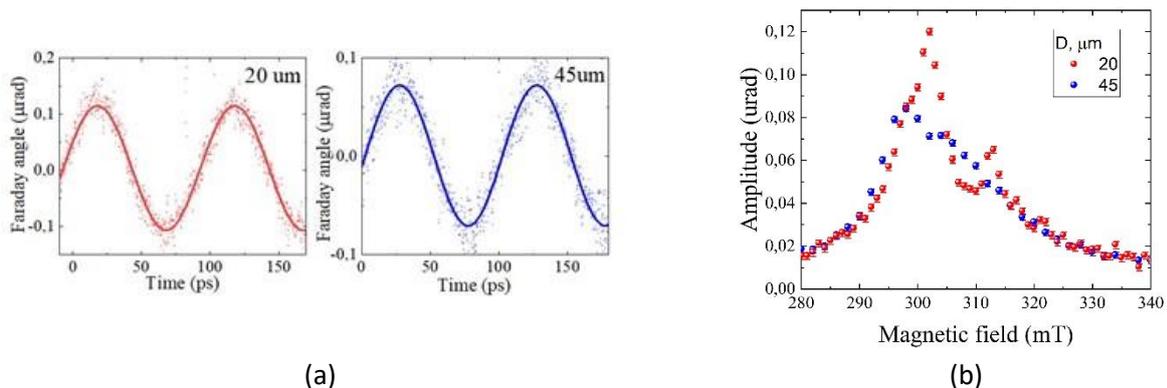

(a)            (b)

**Figure 3.** (a) Magnetization precession excited in the 20 μm and 45 μm disks when the pump beam center coincides with the microdisk center. The external magnetic field is 3,01 kOe. (b) Dependence of the magnetization precession amplitude on the magnetic field for the 20 μm and 45 μm disks. The laser pump fluence is 5,2 μJ/cm$^2$.

The experimental studies were conducted on a 1.1 μm-thick magnetic dielectric film of Bi-substituted rare-earth iron garnet (BiLuGdEu)$_3$Fe$_5$O$_{12}$, grown on a gadolinium gallium garnet substrate with the crystallographic orientation (111). The saturation magnetization is $4\pi M_s$=1300 G, the uniaxial anisotropy constant is $K_u = 3 \cdot 10^3 \frac{\text{erg}}{\text{cm}^3}$, and the Gilbert constant is $\alpha = 4 \cdot 10^{-3}$. is dissipation constant used in the Landau − Lifshitz − Hilbert equation. Microdisks of various diameters were etched in the film down to the substrate using a photolithography and wet etching process (Fig.1(e)).

To study the laser excitation of spin waves in the disks, the sample is placed in a saturating in-plane external magnetic field and a pump-probe method based on asynchronous optical sampling (ASOPS) (Supplementary, Fig. S1) is implemented[37] . The circularly polarized laser pump pulses of duration 50 fs at the wavelength $\lambda_{pump}$=810 nm are focused in a spot of 9 μm in diameter and hit the sample each 100 ps, initiating the spin dynamics, which can be described in a spherical coordinate system by the polar and azimuth angles $\theta_1$ and $\varphi$, respectively, where the z-axis is directed along the film normal (Fig. 3(a)). The spin precession takes place around the external magnetic field directed along the x-axis. The spins remain close to the equilibrium direction along the x-axis and follow an elliptical trajectory so that $\theta = \frac{\pi}{2} - \theta_1 \ll 1$ and $\varphi \ll 1$. The spin precession is monitored via the Faraday effect by the probe beam at $\lambda_{probe}$=780 nm wavelength focused in a spot of 7 μm in diameter. The probe beam arrives at the sample at a variable time delay with respect to the pump which allows to observe the magnetization vector changes with temporal resolution. The rotation of the probe polarization by the Faraday angle $\Phi$ provides the angle $\Phi$: $\theta = \frac{\Phi}{\Phi_0}$, where $\Phi_0$ is the Faraday angle for the magnetic film fully saturated out-of-plane.

*Lett.* **42**, 279–282 (2017).

**ACKNOWLEDGEMENTS**

This work was financially supported by a Russian Foundation of Basic Research (project N 20-52-12047) and Deutsche Forschungsgemeinschaft (project No. AK40/11-1). Growth of the iron garnet films was financially supported by the Russian Ministry of Education and Science, Megagrant project N 075-15-2019-1934. Also support by the DFG within the International Collaborative Research Center TRR 160 (project A1 and C2) is appreciated. Anastasiya Khramova acknowledges the support from BASIS Foundation scholarship (18-2-6-202-1).

V.I. Belotelov is a member the Interdisciplinary "Scientic and Educational School of Moscow University Photonic and Quantum technologies. Digital medicine.".

**AUTHOR CONTRIBUTIONS STATEMENT**

M.K., I.V.S., A.E.K., I.A.A. conducted the experimental part of the study, A.E.K. and V.I.B. prepared the manuscript, A.K.Z., I.A.A., M.B. and V.I.B. conceived the project, M.A.K. was a scientific consultant, A.N.S. and V.N.B. made samples. All authors edited and reviewed the manuscript.

**Supplementary**

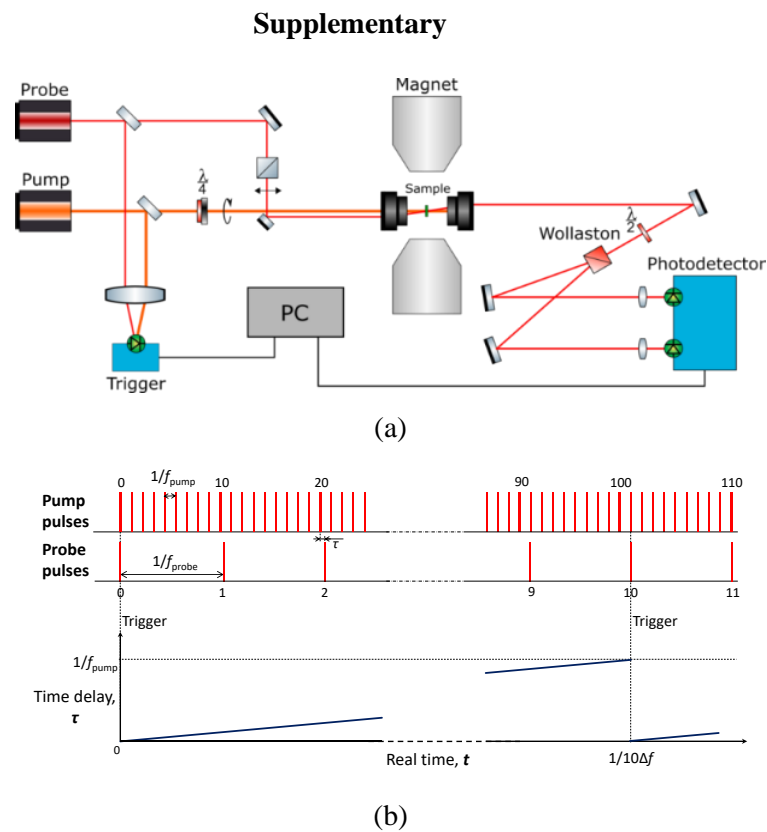

(a)

(b)

FIG. S1. Asynchronous optical sampling (ASOPS) method.

Asynchronous optical sampling (ASOPS) method

The magnetization precession is excited and detected using a pump-probe method based on asynchronous optical sampling (ASOPS). This method involves the operation of two independent Ti: Sapphire femtosecond lasers, which emit pump and probe pulses at the center wavelength of λ ≈ 810 nm and 780 nm. The pump pulses have a repetition rate of $f_{pump}$ = 10 GHz, and the probe pulses have a rate of $f_{probe}$ = $f_{pump}$/10 - Δf, where Δf = 2 kHz (Fig S1a). Thus, each probe pulse is synchronized with every 10th pump pulse with a small offset Δf. As a result, the relative time delay between the pump and the 10th probe pulse periodically increases from zero to 100 ps during a scan time of 50 μs. The fast signal is linearly stretched in time with a factor of about $f_{probe}$/Δf and makes it accessible to fast data acquisition electronics (Fig S1b). The pump and probe beams are focused on the sample using a single reflective microscope objective with a magnification of 15, including 4 sectors through which light can enter and exit, pump beam diameter 9 μm, probe beam diameter 7 μm . The second objective of the microscope is used to collect and collimate the probe beam in transmission geometry. Stray light from pump laser in the detection path was reduced by using interference bandpass filter centered at the probe wavelength.

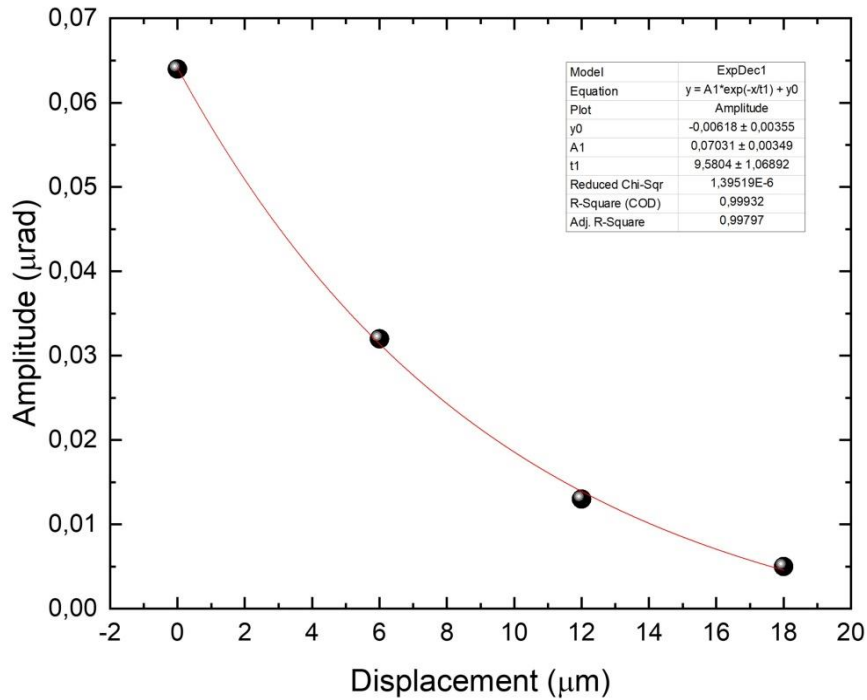

FIG. S2 Exponential decay of spin waves as the sample beam moves away from the pump beam. The black circles indicate the experimental data on the amplitude of the spin waves. The red line is the fit curve. The external magnetic field for this experiment was 2.97 kOe. The pump fluence is 5,2 μJ/cm²